\documentclass[aps,prd,groupedaddress]{revtex4}

\usepackage{amsfonts}
\usepackage{graphicx}

\begin{document}


\title{Proof of triviality of $\lambda\phi^4$ theory}


\author{Marco Frasca}
\email[]{marcofrasca@mclink.it}
\affiliation{Via Erasmo Gattamelata, 3 \\ 00176 Roma (Italy)}


\date{\today}

\begin{abstract}
We show that a recent analysis in the strong coupling limit of the $\lambda\phi^4$
theory proves that this theory is indeed trivial giving in this limit the expansion
of a free quantum field theory. We can get in this way the propagator with the
renormalization constant and the renormalized mass. As expected the theory in this
limit has the same spectrum as a harmonic oscillator. Some comments about triviality
of the Yang-Mills theory in the infrared are also given. 
\end{abstract}


\maketitle


In a series of papers we showed recently that gradient expansions are to be considered as
strong coupling expansions when applied to the study of field theory both classical and
quantum. In general relativity this gives a sound proof of the so called 
Belinski, Khalatnikov and Lifshitz conjecture \cite{fra1}. Analysis of a classical scalar theory
was also given \cite{fra2}. Finally, a quantum theory applied to the symmetric phase
of a single component $\lambda\phi^4$ theory was given \cite{fra3}. Application to the
$N$-component theories was also proposed \cite{fra4}. The main development of these works
arises from a paper about duality principle in perturbation theory \cite{fra5}.

The quantum formulation of the strong coupling expansion given in \cite{fra3} opened up
a problem about an almost completely verified property of the $\lambda\phi^4$ theory, that
is, this theory is trivial in the limit of the coupling going to infinity for 
dimensionality greater than four \cite{zj}. This was proved by Aizenman \cite{aiz}. 
In four dimensions the problem is still open but
results obtained since eighties through some works by L\"uscher and Weisz \cite{lw1,lw2,lw3,lw4} and
Kati and Shen \cite{ks1} seem to give a strong indication that this theory is trivial. 
The present situation in four dimensions is described by Stevenson \cite{ste}
and by Balog, Niedermayer, and Weisz \cite{wei} with respect to a fit of the
propagator requiring more than one term in the free form.
In dimension lesser than four the model is expected to be not trivial. 
Anyhow, it is interesting to note that this property of the scalar theory helped
people to give some limits on the mass of the Higgs boson (see e.g. \cite{ks2}). In this
case a current view is given in \cite{cc1,cc2} analyzing data coming from lattice
computations. 

The point to be emphasized is that one generally assumes that the theory is trivial when
the renormalized coupling constant goes to zero in the limit of a large cut-off and the
theory becomes a free theory in this limit. This limit coincides with the one where the
bare coupling constant goes to infinity, that is the case we discussed analytically in 
\cite{fra3}. Indeed, we would like to consider a redefinition of the triviality of a quantum
field theory in the limit of a bare coupling going to infinity. We define ``trivial'' a
theory that has a two-point function in the form
\begin{equation}
    G(p) = \sum_{n}\frac{Z_n}{p^2+m_n^2-i\epsilon}
\end{equation} 
that is a weighted sum of free propagators being $Z_n$ the weights and $m_n$ what we call the
spectrum of the theory. This will include also the particular case of all $Z_n$ being
zero except one. We will prove in the following that for a $\lambda\phi^4$ theory this
is indeed the case in the limit $\lambda\rightarrow\infty$ and we do the
proof by simply summing the series obtained in \cite{fra3}.  

In our work \cite{fra3} we showed that in the limit $\lambda\rightarrow\infty$ the theory
has the same spectrum as a harmonic oscillator and this is a strong clue that the theory is 
trivial in this limit in the sense given above. 
We will now see that the proof is indeed complete and perturbation theory
is consistent with this conclusion. But before we can go on let
us give a simple definition through path integrals of a trivial theory for the scalar field.
It is well-known that the generating functional for the scalar theory can be written as \cite{zj}
\begin{equation}
     Z[j]=\int[d\phi]e^{\left\{i\int d^Dx\left[
     \frac{1}{2}(\partial\phi)^2-V(\phi)+j\phi
     \right]\right\}}
\end{equation}
and a free massive theory is given by taking $V(\phi)=\frac{1}{2}\mu_0^2\phi^2$. 
This means that the above generating functional can be rewritten immediately in a Gaussian form
\begin{equation}
Z[j]=\exp\left[\frac{i}{2}\int d^Dy_1d^Dy_2\frac{\delta}{\delta j(y_1)}\mu_0^2\delta^D(y_1-y_2)
    \frac{\delta}{\delta j(y_2)}\right]Z_0[j]
\end{equation}
being
\begin{equation}
    Z_0[j]=\exp\left[\frac{i}{2}\int d^Dx_1d^Dx_2j(x_1)\Delta(x_1-x_2)j(x_2)\right]
\end{equation}
with $\Delta(x_1-x_2)$ the propagator of the massless theory. This form of the generating
functional produces immediately the well-known form of the two-point function 
$G(p)=1/(p^2+\mu_0^2-i\epsilon)$. So, whenever we are able to
rewrite a theory in the above form for the generating functional, that is the Gaussian form
\begin{equation}
Z[j]=\exp\left[\frac{i}{2}\int d^Dy_1d^Dy_2\frac{\delta}{\delta j(y_1)}A(x_1)\delta^D(y_1-y_2)
    \frac{\delta}{\delta j(y_2)}\right]Z_0[j]
\end{equation}
with $Z_0[j]$ defined by some two-point function of the kind $\Delta(p)=\sum_{n}Z_n/(p^2+m_n^2-i\epsilon)$
we can immediately write it down the propagator 
in the form  $\Delta(p)=\sum_{n}Z_n/(p^2+\tilde A(p)+m_n^2-i\epsilon)$.
The theory will be proved trivial if the two-point function is indeed a sum of free particle
propagators as defined above.

But this is exactly the case for a scalar theory in the limit $\lambda\rightarrow\infty$
as we have already proved that (assuming $\mu_0=1$) \cite{fra3}
\begin{equation}
    Z[j]=\exp\left[\frac{i}{2}\int d^Dy_1d^Dy_2\frac{\delta}{\delta j(y_1)}(-\nabla^2+1)\delta^D(y_1-y_2)
    \frac{\delta}{\delta j(y_2)}\right]Z_0[j]
\end{equation}
being in this case
\begin{equation}
\label{eq:z0}
    Z_0[j]=\exp\left[\frac{i}{2}\int d^Dx_1d^Dx_2j(x_1)\Delta(x_1-x_2)j(x_2)\right]
\end{equation}
and
\begin{equation}
\label{eq:p1}
    \Delta(\omega)=\sum_{n=0}^\infty\frac{B_n}{\omega^2-\omega_n^2+i\epsilon}
\end{equation}
being $K(i)$ the constant
\begin{equation}
    K(i)=\int_0^{\frac{\pi}{2}}\frac{d\theta}{\sqrt{1+\sin^2\theta}}\approx 1.3111028777,
\end{equation}
and
\begin{equation}
    B_n=(2n+1)\frac{\pi^2}{K^2(i)}\frac{(-1)^{n+1}e^{-(n+\frac{1}{2})\pi}}{1+e^{-(2n+1)\pi}}.
\end{equation}
The mass spectrum of the theory is given by
\begin{equation}
\label{eq:ms}
    \omega_n = \left(n+\frac{1}{2}\right)\frac{\pi}{K(i)}\left(\frac{\lambda}{2}\right)^{\frac{1}{4}}
\end{equation}
proper to a harmonic oscillator. Now, let us see why in lattice
computations the form of the propagator appears to be \cite{ks1}
\begin{equation}
\label{eq:p2}
    \Delta(p)=\frac{Z_R}{p^2+m_R^2}
\end{equation}
being $Z_R$ a renormalization constant and $m_R$ the renormalized mass. Indeed, higher
order corrections in our expansion gives the missing part between the propagator
(\ref{eq:p1}) and (\ref{eq:p2}) in the momentum as we showed above. But one has
\begin{equation}
    Z_R=B_0=\frac{\pi^2}{K^2(i)}\frac{e^{-\frac{\pi}{2}}}{1+e^{-\pi}}\approx 1.144231098
\end{equation} 
of the order of unit as it should be \cite{ks1}. The next terms are exponentially damped, e.g.
\begin{equation}
    Z'_R=B_1=\frac{3\pi^2}{K^2(i)}\frac{e^{-\frac{3\pi}{2}}}{1+e^{-3\pi}}\approx .1483401274
\end{equation} 
smaller by one magnitude order due to e-folding. The renormalized mass,
here given by $\frac{\pi}{2K(i)}\left(\frac{\lambda}{2}\right)^{\frac{1}{4}}$
for $\lambda$ large enough, contributes
to damp out higher order terms as it increases with the square of the order 
making these terms practically not visible on the lattice. So,
it would be really interesting to do some computer work to measure them and see all the
spectrum of the theory in this limit. We just notice that a strong indication that a further
term should be added to the free propagator for the scalar theory in four dimensions, in
order to improve the fit to numerical data, has been given in \cite{ste,wei}. This further term
has a mass being three times the mass of the first term in the symmetric phase 
in fully agreement with our analysis.

In order to see how our perturbation theory produces terms proper to a trivial theory we push
the analysis to second order for the computation of the generating functional. As we will
see, no infinite terms appear, rather one gets only the proper momentum contribution. In
Ref.\cite{fra3} we obtained
\begin{equation}
    Z[j]=\left[1-\frac{i}{2}\int
	\int d^Dy_1(-\nabla^2_{y_1}+1)\int d^Dx_1d^Dx_2\Delta(y_1-x_1)\Delta(y_1-x_2)j(x_1)j(x_2)
	+\ldots
	\right]Z_0[j].
\end{equation}
Now we have to compute the term
\begin{eqnarray}
     Z^{(2)}[j]&=&-\frac{1}{8}\int d^Dy_1d^D_2\frac{\delta}{\delta j(y_1)}(-\nabla^2+1)
	 \delta^D(y_1-y_2)\frac{\delta}{\delta j(y_2)}
	\times \\ \nonumber
	 & & \int d^Dy_3d^Dy_4\frac{\delta}{\delta j(y_3)}(-\nabla^2+1)
	 \delta^D(y_3-y_4)\frac{\delta}{\delta j(y_4)}Z_0[j]
\end{eqnarray}
being $Z_0[j]$ given by eq.(\ref{eq:z0}). The computation is very easy to be carried out to give
\begin{eqnarray}
    Z^{(2)}[j]&=&\frac{1}{8}\int d^Dy_1d^Dy_2(-\nabla^2_{y_1}+1)(-\nabla^2_{y_2}+1)\times \\ \nonumber
	& &\left[i\int d^Dx_2d^Dx_3\Delta(y_2-y_1)\Delta(y_2-x_2)\Delta(y_1-x_3)j(x_2)j(x_3)\right. \\ \nonumber
	&+&i\int d^Dx_1d^Dx_3\Delta(y_2-x_1)\Delta(y_2-y_1)\Delta(y_1-x_3)j(x_1)j(x_3) \\ \nonumber
	&+&2i\int d^Dx_1d^Dx_2\Delta(y_2-y_1)\Delta(y_2-x_1)\Delta(y_1-x_2)j(x_1)j(x_2) \\ \nonumber
	&-&\left.\int d^Dx_1d^Dx_2d^Dx_3d^Dx_4\Delta(y_2-x_1)\Delta(y_2-x_2)\Delta(y_1-x_3)
	\Delta(y_1-x_4)j(x_1)j(x_2)j(x_3)j(x_4)\right]\times \\ \nonumber
	& &Z_0[j].
\end{eqnarray}  
Doing this computation a term appears having the form
\begin{equation}
    \frac{1}{4}\int d^Dy_1d^Dy_2[\Delta(y_2-y_1)]^2
\end{equation}
that seems to diverge. Indeed, when we introduce a cut-off in momentum and take the limit
of the cut-off going to infinity, this term is easily proven to be zero. So, we can conclude
that our perturbation theory just produces the missing contribution in momentum to the
propagator proper to a free theory and the theory is finally shown to be trivial in the
limit of the coupling going to infinity. The value of this result relies also on the
fact that results from weak perturbation theory that are used for renormalization group
analysis, e.g. to compute the beta function in charge renormalization, can also be trusted
and all the picture by both sides of perturbation theory appears fully consistent. So, the
two-point function of a scalar theory in the limit of the coupling going to infinity
has just the simple form of a weighted sum of free propagators with a mass spectrum being the
same of a harmonic oscillator.

Finally, we would like to spend a few words about Yang-Mills theory. It is well-known that
on a compact manifold the theory has the renormalized coupling going to zero in the infrared
\cite{cf1}. This result has been reently confirmed notwithstanding the change into the
behavior of the propagator \cite{cfn}. 
This result seems to translate to the theory on a non-compact manifold
through some lattice computations while disagrees with Dyson-Schwinger equation analysis
\cite{ste2}. So, let us assume as a working hypothesis that both the results on
a compact manifold and lattice computations about the running coupling are the one to be trusted. 
We can conclude that Yang-Mills theory is trivial, in the sense given above, in the infrared limit. 
This hypothesis has some interesting consequences. Firstly, one can immediately write down the 
propagator for the gluon
\begin{equation}
    D(p)=\sum_{n=0}^\infty\frac{Z_n}{p^2+m_n^2}
\end{equation}
being $Z_n$ renormalization constants to be computed, 
$m_n$ the $0^{++}$ glueball mass spectrum that should go as the one of a harmonic oscillator
\begin{equation}
\label{eq:sp}
    m_n=\left(n+\frac{1}{2}\right)\delta
\end{equation}
and $\delta/2$ the mass gap of the theory. This in turn will imply that the propagator is finite
in the limit of the momentum going to zero and that the scaling law in the infrared for
the propagator has the exponent $\kappa=0.5$. Besides, computations on a lattice for
the case of 2+1 dimensions also agree with our analysis of a harmonic oscillator
spectrum for Yang-Mills theory \cite{mck1,mck2}. A full accounting of this situation has been
presented in \cite{fra4}.

In order to analyze further this hypothesis about the Yang-Mills theory, we compare our propagator
to the one obtained in the lattice \cite{ste2}. To make sense out of this comparison the agreement
must be reached after we fix the only free parameter, that is the factor $\delta/2$ into eq.(\ref{eq:sp}).
This is the gluon mass. So we take \cite{fra4}
\begin{equation}
    Z_n=(2n+1)\frac{\pi^2}{K^2(i)}\frac{(-1)^{n+1}e^{-(n+\frac{1}{2})\pi}}{1+e^{-(2n+1)\pi}}
\end{equation}
and $m_n=(2n+1)m_g$ being $m_g$ the gluon mass to be fitted. 
\begin{figure}[tbp]
\begin{center}
\includegraphics[angle=0,width=240pt]{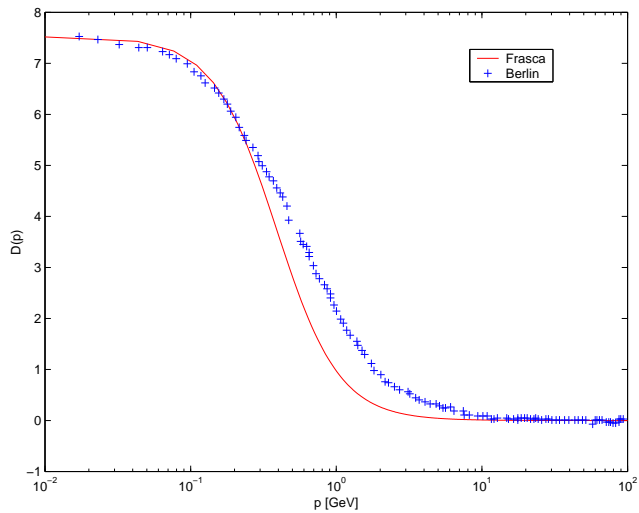}
\caption{\label{fig:fig1} Comparison between lattice gluon propagator and our analytical result.}
\end{center}
\end{figure}
The result is given in fig.(\ref{fig:fig1}) for $m_g=389$ MeV. The agreement is excellent supporting
our conclusions.

We proved that our approach for strongly coupled quantum field theory applied to a
scalar field permits to prove that this theory is trivial in the strong coupling limit.
Similar considerations applied to Yang-Mills theory in the infrared permit to obtain some interesting
results about the gluon propagator and the mass spectrum of the theory. In this latter case the
comparison with lattice results is excellent for a gluon mass of 389 MeV. 

\begin{acknowledgments}
I have to thank Jean Zinn-Justin for a clarifying communication about the main argument of
this paper.
\end{acknowledgments}

\end{document}